\documentclass[useAMS,usenatbib]{mn2e}
\usepackage{graphicx}
\usepackage{amsmath}
\usepackage{amssymb}
\usepackage[draft]{hyperref}
\usepackage{subcaption}
\newcommand{\poubelle}[1]{}

\usepackage{lineno,xcolor}

\title[On the search for Galactic supernova remnant PeVatrons with current TeV instruments]{On the search for Galactic supernova remnant PeVatrons with current TeV instruments}
\author[P. Cristofari et al.]
{P. Cristofari$^{1}$\thanks{pc2781@columbia.edu},
S. Gabici$^{2}$,
R. Terrier$^{2}$, 
T. B. Humensky$^{3}$
\\
$^{1}$Department of Astronomy, Columbia University, 10027 New York, USA\\
$^{2}$APC, AstroParticule et Cosmologie, Universit\'e Paris Diderot, CNRS/IN2P3, CEA/Irfu, Observatoire de Paris, Sorbonne Paris Cit\'e,\\ 10, rue Alice Domon et L\'eonie Duquet, 75205 Paris Cedex 13, France \\
$^{3}$Physics Department, Columbia University, 10027 New York, USA
}

\begin{document}

\date{}

\pagerange{\pageref{firstpage}--\pageref{lastpage}} \pubyear{}

\AtEndDocument{\label{lastpage}}

\maketitle

\label{firstpage}

\begin{abstract}
The supernova remnant hypothesis for the origin of Galactic cosmic rays has passed several tests, but the firm identification of a supernova remnant  pevatron, considered to be a decisive step to prove the hypothesis, is still missing. While a lot of hope has been placed in next--generation instruments operating in the multi--TeV range, it is possible that current gamma--ray instruments, operating in the TeV range, could pinpoint these objects or, most likely, identify a number of promising targets for instruments of next generation. Starting from the assumption that supernova remnants are indeed the sources of Galactic cosmic rays, and therefore must be pevatrons for some fraction of their lifetime, we investigate the ability of current instruments to detect such objects, or to identify the most promising candidates.  
\end{abstract}


\section{Introduction}
\label{sec:introduction}
Supernova remnants (SNRs) are by far viewed as the most probable sources of  Galactic cosmic rays (CRs). 
The requirements for a potential candidate to be the source of Galactic CRs are imposed by CR measurements. Amongst the most essential constraints imposed by this hypothesis, one can mention four of them: the observed level of CRs at the Earth, the CR lifetime in the Galaxy, the remarkably broad power--law energy spectrum, and the extension of this power--law spectrum  up to an energy of $\sim$ PeV, called the \textit{knee}~\citep{antoni2005,bartoli2015}, where the power--law spectrum breaks. 
Thus, if SNRs are the sources of Galactic CRs, they have to be able to meet these four essential requirements. 

The observed intensity of CRs and their estimated lifetime in the Galaxy are compatible with the SNR hypothesis, provided that a fraction $\approx $ 10\% of the total explosion energy of the supernova progenitor is converted into CRs. Moreover, the observed power--law spectrum of CRs can be explained by the Diffusive Shock Acceleration mechanisms~\citep{ostriker1978,bell1978}. 

The last requirement, i.e., imposing that SNRs must be able to accelerate particles up to the \textit{knee}, can also be met provided that efficient magnetic field amplification happens at SNR shocks~\citep{bell04,drury2012}. For more references, see e.g. \citet{gabici2016}. In this hypothesis, SNRs thus have to be \textit{pevatrons} for some time in their lifetime. 
At this stage, no detection of a SNR pevatron has been reported, but the search for such objects is strongly motivated as it would  constitute major evidence in favor of the SNR hypothesis. 

The recent detection of a pevatron located in the galactic centre by the H.E.S.S. Collaboration~\citep{HESSpevatron} triggered a widespread discussion in the community \citep[see e.g.][]{gaggero2017,terrier2017}.
Remarkably, it demonstrates the feasibility of pevatron searches in the TeV sky, and suggests that sources other than SNRs might also accelerate PeV CRs in the Galaxy.

The observations of several SNRs in the TeV gamma--ray range have provided evidence that efficient particle acceleration is happening at SNRs shocks. Indeed, the production of gamma rays in this range can be explained by two types of interactions. One the one hand, accelerated protons can interact with protons of the interstellar medium, and these hadronic interactions can produce gamma rays through pion decay. On the other hand, accelerated electrons can scatter off photons of the Cosmic Microwave Background, or other soft photon fields from stellar emission or dust, and these leptonic interactions, referred to as Inverse Compton Scattering, can produce gamma rays. 
Because of these two competing mechanisms, it is in most cases difficult to draw firm conclusions regarding the mechanism involved in the production of  gamma rays in the TeV range~\citep[see e.g.][and references therein]{gaggero2018}. 

It is, however, noteworthy that in the multi--TeV energy domain, the situation becomes clearer. Because of the Klein--Nishina suppression of the cross section for inverse Compton scattering, the production of gamma rays via leptonic mechanisms becomes insignificant at energies above tens of TeV. Therefore, the detection of a gamma-ray spectrum extending without any attenuation up to the multi--TeV range constitutes a proof for hadronic interactions \citep[e.g.][]{gabici2007,HESSpevatron}. 


In light of this fact, we propose in this article a study of the expected characteristics (e.g. flux, spectrum, angular size, etc.) that SNR pevatrons would exhibit when observed in the TeV and multi--TeV range by current instruments. 
Starting from the assumption that SNRs are the sources of Galactic CRs, and using Monte--Carlo simulations, we perform a population study for these objects. 
This will then give the expected number of SNR PeVatrons among the sources detected in the H.E.S.S. Galactic Plane Survey \citep{donath2016}.
We will also discuss the role of multi-wavelength observations (especially in the X-ray band), and of present and future facilities such as HAWC (\url{www.hawc-observatory.org}) and the Cherenkov Telescope Array (CTA, \url{www.cta-observatory.org}).

It has to be stressed that the detection of a SNR PeVatron by a given telescope operating in the TeV or multi-TeV domain would not necessarily imply that the detected object is indeed recognized as a PeVatron. This is because the low photon statistics in the multi-TeV domain might prevent an accurate reconstruction of the shape of the parent CR spectrum up to PeV particle energies. Therefore the main goal of this paper is to identify the observational properties that a SNR PeVatron would exhibit when observed by currently operating gamma-ray instruments.


In this work, we rely on the approach presented in~\citet{cristofari1,cristofari2}. In Sec.~\ref{sec:SNRs}, we briefly describe the approach and the differences with the previously cited articles. Our results are presented in Sec.~\ref{sec:results} and we summarize our work in Sec.~\ref{sec:conclusions}. 

\section{A Monte Carlo approach}
\label{sec:SNRs}
The procedure used in this work was presented in~\citet{cristofari1,cristofari2}. We refer the reader to these papers for a more detailed description. 
In this approach, we begin by simulating the age and location of supernovae in the Galaxy, assuming a typical rate $\nu_{\rm SN} = 3 /$century~\citep[see e.g.][and references therein]{li2011} and a spatial distribution based on the description of~\citet{PSR} and \citet{lorimer}. 

The evolution of each simulated SNR is then computed following the description provided by~\citet{chevalier}, \citet{pz05}, \citet{ostriker}, and \citet{bisnovati}. Following the approach of~\citet{pz05}, we then compute the spectrum of accelerated protons and electrons, and finally the gamma--ray luminosity from each SNR shell \citep{cristofari1,cristofari2}. The gas density at the location of the simulated SNRs is taken from \citet{H1} and \citet{H2} for atomic and molecular hydrogen, respectively. The two crucial assumptions made here are: {\it i)} protons and electrons are accelerated at the shock with a power--law spectrum $f(p) \propto p^{-\alpha}$, with $p$ the momentum of a given particle, {\it ii)} a fraction $\xi_{\rm CR} \sim 0.1$ of the shock ram pressure is transferred into CRs. We computed $10^3$ realizations of the Galaxy, a number sufficient to study the average properties of the pevatron population. 
In the following, two Types of SN progenitors are considered:~thermonuclear (Type Ia) with a relative rate of appearance of 0.32 and core--collapse (all other Types) with relative rate 0.68~\citep[see e.g.][]{seo}. The typical parameters assumed for these two Types of progenitors are summarized in Table~\ref{tab:Types}. 

\begin{figure}
\includegraphics[width=.5\textwidth]{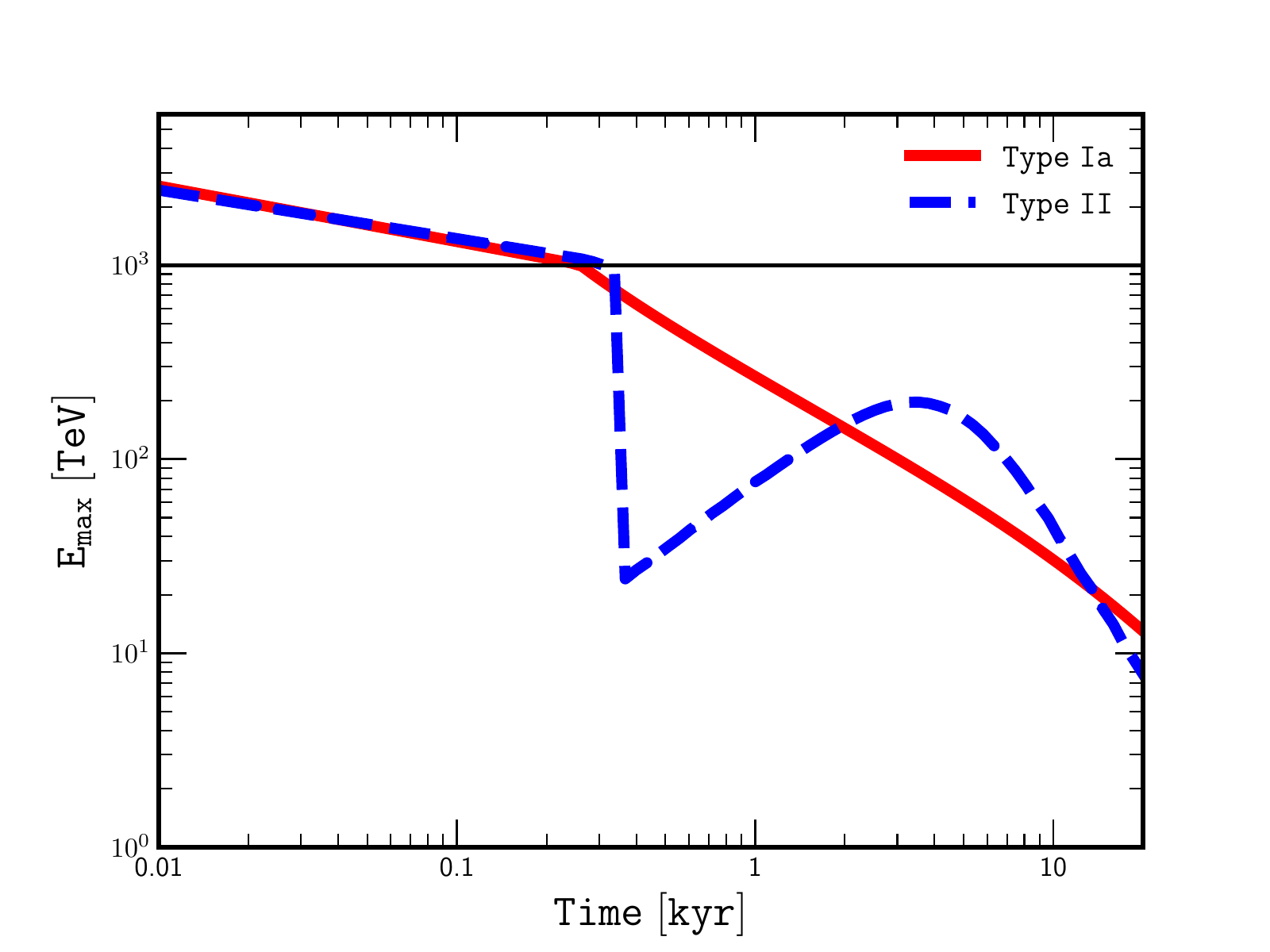}
\caption{Maximum energy of particles accelerated at the SNR shock. The red (solid) line corresponds to Type Ia progenitors and the blue (dashed) line corresponds to Type II progenitors. In this example, the transition between the FEP and SP is at $\approx 0.26$ kyr and $\approx 0.3$ kyr for Type Ia and Type II respectively where an energy of 1~PeV is reached. A density $n_{0}=1$~cm$^{-3}$ is assumed for the ISM. The horizontal black line shows the 1~PeV threshold.}
\label{fig:Emax}
\end{figure}

A central assumption done in this paper is that all SNRs are pevatrons at some stage of their evolution and that they are the sources of the CRs observed in the region of the knee, featured at an energy $E_{\rm knee}\approx 1$ PeV. In order to reproduce the knee feature observed in the CR spectrum we have to impose that SNRs accelerate protons up to the energy $E_{\rm knee}$ at the transition between the free expansion and Sedov phase of their evolution, as illustrated in Fig.~\ref{fig:Emax}.
Such a scenario challenges current theoretical models~\citep[see e.g.][]{schure2013}, and is not easily testable by means of gamma-ray observations, mainly due to the poor photon statistics obtained in gamma-ray observations of SNRs in the multi-TeV band.
For these reasons, our approach is purely phenomenological, and in the following we simply {\it impose} that SNRs can explain the CRs observed at the energy of the knee: CR protons of energy $E_{knee}$ escape the SNR at the transition between the free-expansion and Sedov phases.  

The evolution of the maximum energy of accelerated particles at a SNR shock is computed assuming that particles escape once their diffusion length equals a fraction $\zeta_{esc}$ of the shock radius, i.e. $E_{\rm max} \propto u_{\rm sh} B R_{\rm sh}$, where B is the magnetic field at the shock, $u_{\rm sh}$ and $R_{\rm sh}$ the velocity and radius of the shock. Imposing 1 PeV at the transition between SP and EP amounts to making the assumption of efficient magnetic field amplification at the shock.
For the interstellar medium parameters reported in Table~\ref{tab:Types} and for $E_{knee} = 1$ PeV (see Fig.~\ref{fig:Emax}), the magnetic field upstream of the shock at the transition between FE and SP reaches $\approx 100 ~(\zeta_{esc}/0.1)^{-1} \mu \text{G}$ and $\approx 220~(\zeta_{esc}/0.1)^{-1} \mu \text{G}$ for Type Ia and II respectively.
These somewhat extreme values reflect the fact that it is difficult to accelerate PeV particles at SNR shocks, but they are indeed required to boost the energy of CR protons up to the PeV domain.

Different measurements of the position of the knee in the proton spoectrum of CRs have pointed to values in the range [700 TeV -- 3 PeV]~\citep{antoni2005,bartoli2015}. We will further discuss the implications that values in this range might have.

 \begin{table}
\centering
\begin{tabular}{c c c c c c}
\hline 
Type & ${\cal E}_{51}$ & $M_{ej,\odot}$ & $\dot{M}_{-5}$ & $u_{w,6}$ & Rel. rate\\
\hline 
\hline
Ia & 1 & 1.4  & -- & -- & 0.32\\
II & 1 & 5 & 1 & 1  & 0.68 \\
\hline
\end{tabular}
\caption{Supernova parameters adopted: progenitor Type, explosion energy in units of $10^{51}$~erg, mass of ejecta in solar masses,  wind mass loss rate in $M_{\odot}$/yr, wind speed in units of 10~km/s, and relative explosion rate. Values from \citet{seo}.}
\label{tab:Types}
\end{table}

\section{Results \& Discussion}
\label{sec:results}
We adopt the definition of \textit{pevatron} for a SNR accelerating PeV particles, i.e. $E_{\rm max} \geq 1$ PeV, and will investigate the number of pevatron detectable by current generation instruments, though, as stressed in Sec.\ref{sec:introduction}, their detection does not directly imply that they can be identified as pevatron.

\begin{figure}
\includegraphics[width=.5\textwidth]{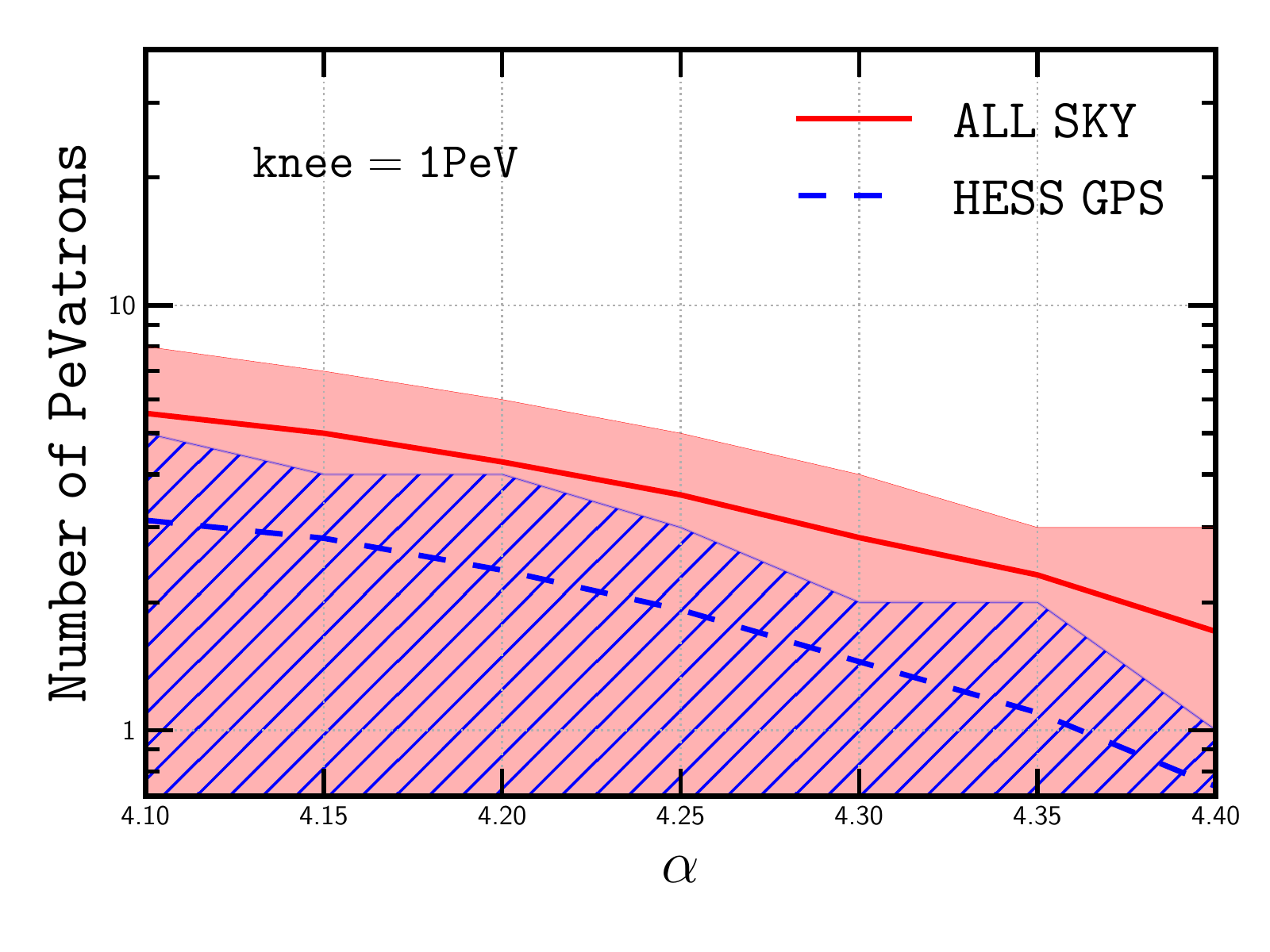}
\caption{Number of pevatrons with integral flux above 1 TeV greater than 1\% of the Crab, as a function of slope of particles accelerated at the shock $\alpha$. The red (solid) line corresponds to the entire sky. The blue (dashed) line corresponds to a region~$ 260^{\circ}<  l < 70^{\circ}$, $| b | < 3^{\circ}$. The shaded area corresponds to +/- 1 standard deviation to the mean. We assume a maximum energy reaching 1~PeV at the transition between the SP and FEP.}
\label{fig:alpha_plot}
\end{figure}

We start by assuming that the position of the knee is $1$ PeV and by computing the number of simulated pevatrons with an integral flux for photons of energy greater than 1 TeV above 1\% that of the Crab Nebula~\citep{2006crab}. This sensitivity is typically reached for pointed observations in $\approx 25$ hours with H.E.S.S.~\citep{2006crab}, and here is taken as a reference value for the performances of current gamma-ray telescopes in the TeV range. In Fig.~\ref{fig:alpha_plot}, the number of pevatrons is plotted as a function of $\alpha$ (the assumed slope of the CR spectrum at SNR shocks) for the entire sky and for a region comparable to the one observed during the H.E.S.S. Galactic plane survey:~$ 260^{\circ}<  l < 70^{\circ}$, $| b | < 3^{\circ}$~\citep{donath2016}.
For hard spectra ($\alpha=4.1$) the mean number of simulated pevatrons detectable by the H.E.S.S. GPS is $3.1$
, for $5.5$
in the entire sky. For steeper spectra $\alpha=4.4$, the number of detections is $1.7$
and $0.7$
for the H.E.S.S. GPS and the entire sky, respectively. 
However, within one standard deviation (shaded regions in Fig.~\ref{fig:alpha_plot}) results are compatible with no detections. In this sense, for a knee at 1 PeV, our results account for the possibility of detecting a few pevatrons, or none, depending simply on the slope of particles accelerated at the shock, and on large fluctuations between different realizations of the Galaxy. In other words, even in the optimistic scenario where we hypothesize that every SNR accelerates PeV particle at the transition between the SP and the FEP, there is a reasonable situation in which no pevatron can be detected during the H.E.S.S. GPS.

\begin{figure}
\includegraphics[width=.5\textwidth]{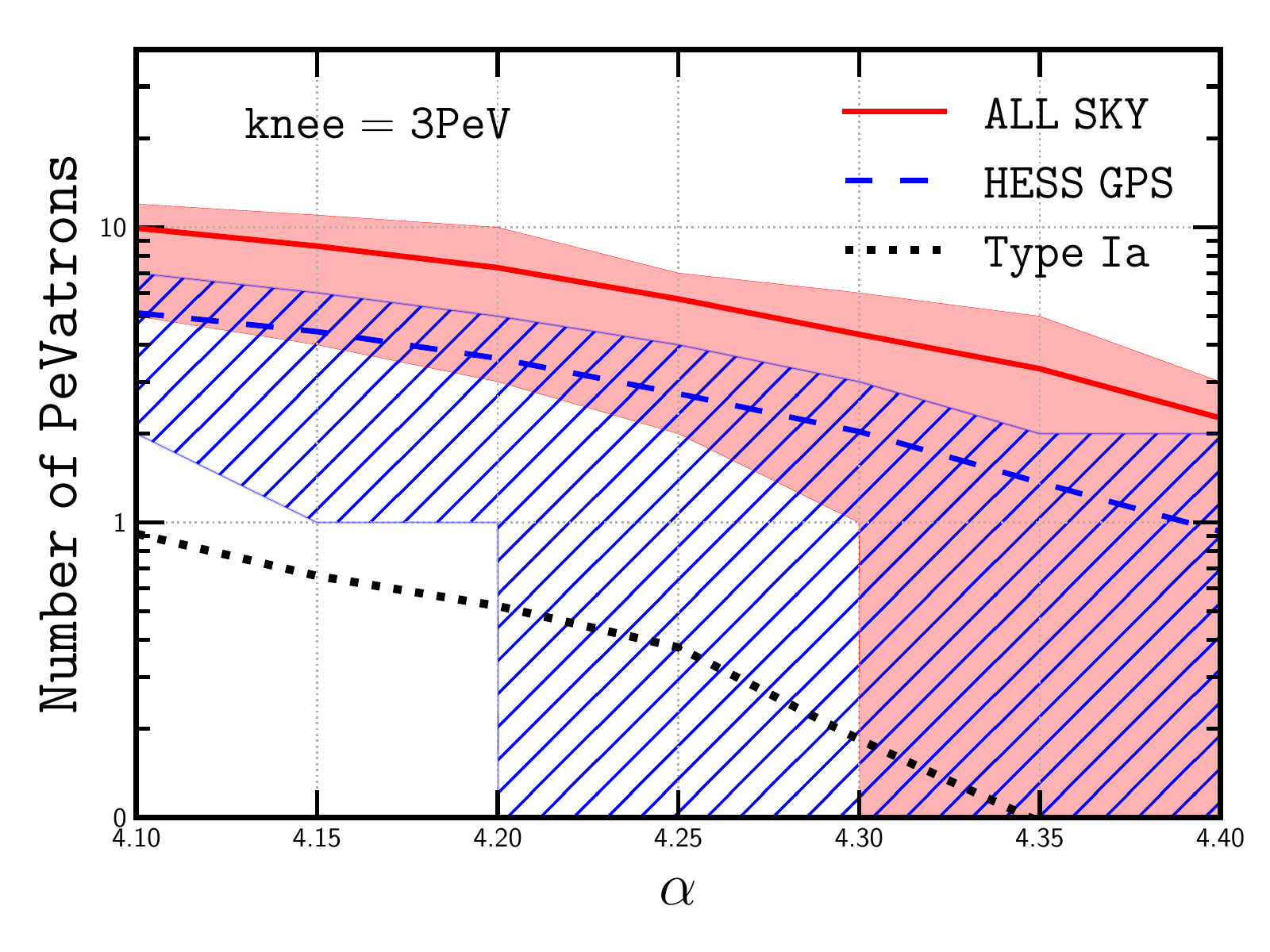}
\caption{Number of pevatrons with integral flux above 1 TeV greater than 1\% of the Crab (green solid curve), as a function of slope of particles accelerated at the shock $\alpha$. The red (solid) line corresponds to the entire sky. The blue (dashed) line corresponds to a region~$ 260^{\circ}<  l < 70^{\circ}$, $| b | < 3^{\circ}$. The shaded area corresponds to +/- 1 standard deviation to the mean. We assume a maximum energy reaching 1~PeV at the transition between the SP and FEP. The number of Type Ia progenitors is represented (black dotted line, HESS GPS).}
\label{fig:alpha2}
\end{figure}

Much more promising results are obtained if the value of $E_{\rm max}$ at the transition between the FEP and the SP is 3~PeV (green solid curve in Fig.~\ref{fig:alpha2}). This value is in agreement with the position of the knee reported by the KASCADE Collaboration \citep{antoni2005}. 
The mean number of detections in the H.E.S.S. GPS goes from $\approx 4.8^{+2.2}_{-4.8}$ to $\approx 1^{+1}_{-1}$ for $\alpha$ in the range [4.1 - 4.4]. For the entire sky, these number become $\approx 9.9^{+2.1}_{-4.9}$ and $\approx 2.3^{+0.7}_{-2.3}$, respectively. 


Regardless of the value of $\alpha$, the majority of the simulated pevatrons detectable in TeV gamma rays come from core--collapse progenitors ($\geq 80$\%), as illustrated by the black dotted line of Fig.~\ref{fig:alpha2} (referring to the H.E.S.S. GPS). This can be explained by two elements. First, Type II supernovae are the most abundant class (we assume here that $\approx 70\%$ of SNR come from Type II supernovae). Second, shocks from core--collapse supernovae expands in a structured medium: starting in a dense wind, they then reach a low density cavity inflated by the wind~\citep{weaver1977}. The density of the unperturbed ISM fixes the density of the low-density bubble, therefore fixing the typical radius at which the transition from the dense wind to the low-density bubble happens. For the set of parameters assumed in this paper, SNRs from Type II progenitors which are in the pevatron phase are all young enough to be still contained within the dense wind. The wind density is typically scaling as~\citep{pz05}: 
\begin{equation}
n_{\rm wind} \approx 2.2 \; \frac{\dot{M}_{-5}}{u_{\rm w,6}} \left(\frac{R}{\text{parsec}} \right)^{-2}\text{cm}^{-3},
\label{eq:nw}
\end{equation}
where $u_{\rm w,6}$ is the wind velocity in units of $10^{6} \text{cm/s}$,  $\dot{M}_{-5}$ the mass loss rate in units of $10^{-5}$~M$_{\odot}$/yr and $R$ the radius. 
This wind density profiles extends up to a radius $r_{\rm w}$, where the density becomes significantly lower, of the order $\sim 10^{-2}$~cm$^{-3}$. $r_{\rm w}$ can be estimated by equating the wind ram pressure to the thermal pressure of the bubble interior~\citep[e.g.][]{cristofari2}:
\begin{equation}
r_{\rm w} \approx 0.9 \; \left(\dot{M}_{-5}u_{\rm w,6}\right)^{1/2} \left( \frac{n_{0}}{ \text{cm}^{-3}} \right)^{-19/35} \text{pc}
\end{equation}

Eq.~\ref{eq:nw} shows that the density profile of the wind does not depend on the location or the associated ambient ISM density, whereas for a Type Ia progenitor, the ambient density $n_{0}$ changes with the Galactic location and typically takes values in the range $\sim 0.1- 1$~\text{cm}$^{-3}$.
Because the gamma--ray luminosity is directly proportional to the density, we can therefore understand why the Type II population is dominant. 

\begin{figure*}
\begin{center}
 \includegraphics[scale=0.64]{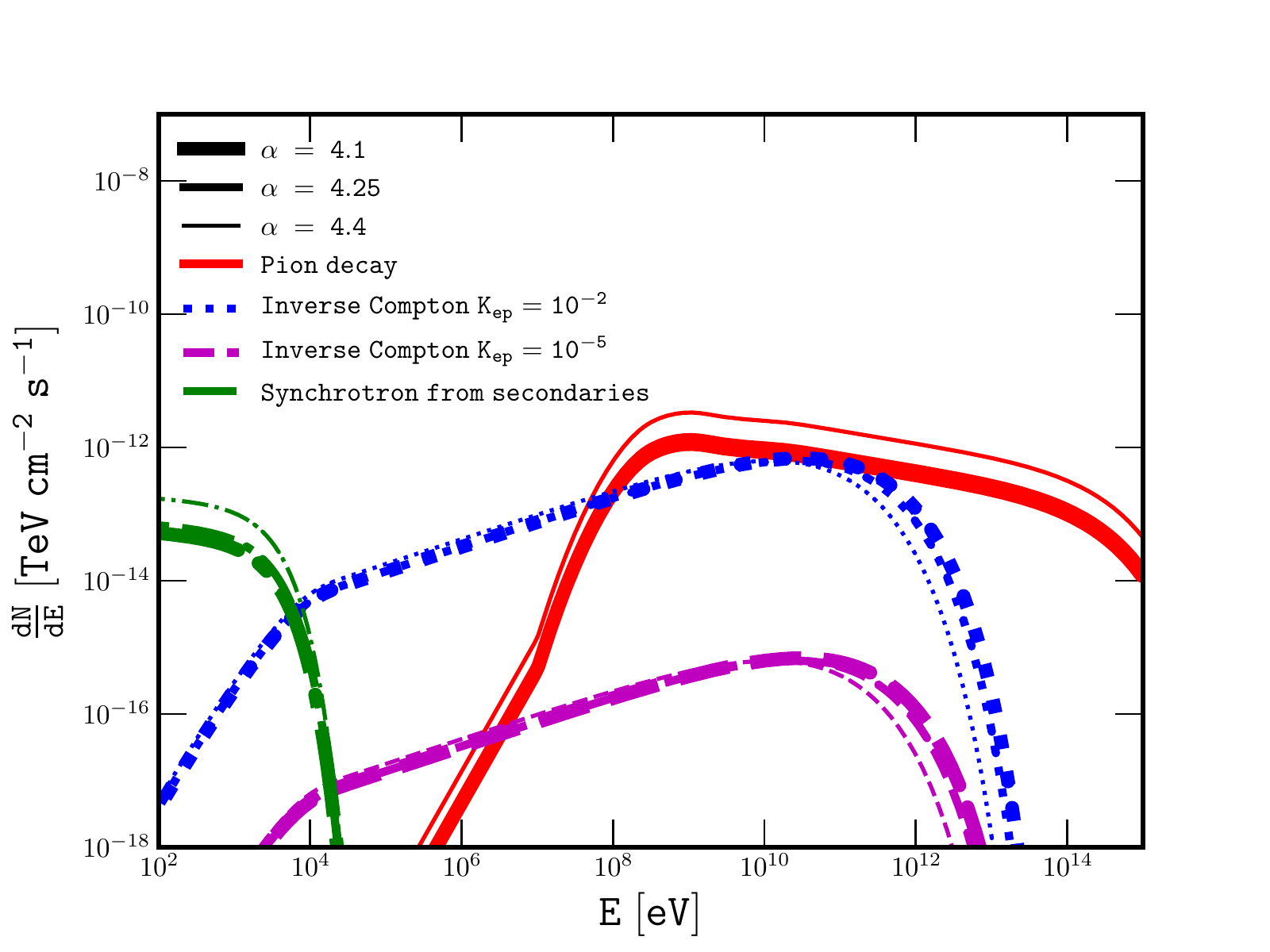}
\caption{Differential spectra of a typical  pevatron, from Type II progenitor, of age $\approx 0.2$ kyr and located at a distance $\approx 7$ kpc. $E_{\rm max}$ at the transition between the SP and the FEP is fixed at 3 PeV. $K_{\rm ep}= 10^{-2}$. The thick, medium, and thin lines correspond to $\alpha$ equal to 4.1, 4.25 and 4.4, respectively.}
\label{fig:spectra}
\end{center}
\end{figure*}

For each simulated pevatron, we can compute the gamma-ray spectrum from pion decay and inverse Compton scattering of primary electrons accelerated at the shock, as described in~\citet{cristofari1,cristofari2}. 
In Fig.~\ref{fig:spectra}, the typical spectrum of a SNR pevatron is presented for a type II supernova progenitor, $E_{knee}$ = 3~PeV, and for a distance of 7 kpc. We consider an electron--to--proton ratio $K_{\rm ep}= 10^{-2}$ (upper dashed curves) and $K_{ep} = 10^{-5}$ (lower dashed curves). 
For $K_{\rm ep}= 10^{-2}$ we remark that in the TeV range the average fluxes from inverse Compton scattering and from pion decay are of the same order, while for $K_{ep} = 10^{-5}$ the hadronic emission largely dominates. The uncertainty in the determination of the parameter $K_{ep}$ implies that it is in general not trivial to ascribe the observed gamma-ray emission to hadronic or leptonic interactions. 
Remarkably, in the multi--TeV range the situation becomes unambiguous. This is because, independently on the value of $K_{\rm ep}$, the gamma-ray emission is always largely dominated by the hadronic contribution. 
A change in the exact value of $E_{\rm max}$ at the transition between the SP and FE phase do not significantly affects these considerations, as long as $E_{\rm max}$ remain within the PeV energy domain.

We additionally compute the spectrum of the population of secondary electrons and positrons produced from proton--proton interactions (charged pion decay), and compute the associated synchrotron emission. Secondary electrons and positrons are computed following the approach of~\citet{kelner}, and their spectrum is obtained under the assumption of fast cooling. 
The fast cooling regime is only valid as long as the synchrotron loss time is shorter than the age of the SNR. We can estimate the minimum energy of electrons satisfying this condition $E_{\rm e,min}$ by equating the age of the considered SNR ($\sim 0.2$ kyr for the SNR represented in Fig.~\ref{fig:spectra}) to the synchrotron loss time: 
\begin{equation}
\tau_{sync}\approx 1.3 \left(\frac{B}{\text{100}\mu \text{G}} \right)^{-2} \left( \frac{E}{\text{TeV}}\right)^{-1} \text{kyr}
\label{eq:sync}
\end{equation}
where E is the energy of the secondaries and B the magnetic field. 
The energy of synchrotron photons emitted scales as $\propto B \times E_{\rm e}^2$, and we can estimate the minimum energy of synchrotron photons $E_{\rm sync, min}$: 
\begin{equation} 
E_{\rm sync, min}\approx 1.7 \left(\frac{t_{\rm age}}{\text{kyr}} \right)^{-2} \left(\frac{B}{100 \mu\text{G}} \right)^{-3} \text{eV}
\end{equation}
For the SNRs considered in our work, of typical age $t_{\rm age}\approx 0.1$~kyr and magnetic field $\sim 100 \mu$G, $E_{\rm sync, min}\approx 0.17$~keV, illustrating that our approach is typically only valid in the entire X-ray donaim. 
The synchrotron emission from secondaries corresponds to the lowest possible X--ray  flux expected by a SNR pevatron.  
Thus, a SNR pevatron whose gamma-ray emission is within the reach of H.E.S.S. should also unavoidably exhibit an X-ray flux roughly at the level of $\gtrsim 10^{-14}$ erg/cm$^2$/s. In fast cooling regime, the ratio between the gamma-ray hadronic emission and the X-ray synchrotron emission from secondary electrons is a constant, and thus brigther gamma-ray SNR pevatrons would also emit a proportionally larger X-ray synchrotron flux from secondaries.


\begin{figure*}
\begin{center}
\minipage{0.48\textwidth}
 \includegraphics[scale=0.56]{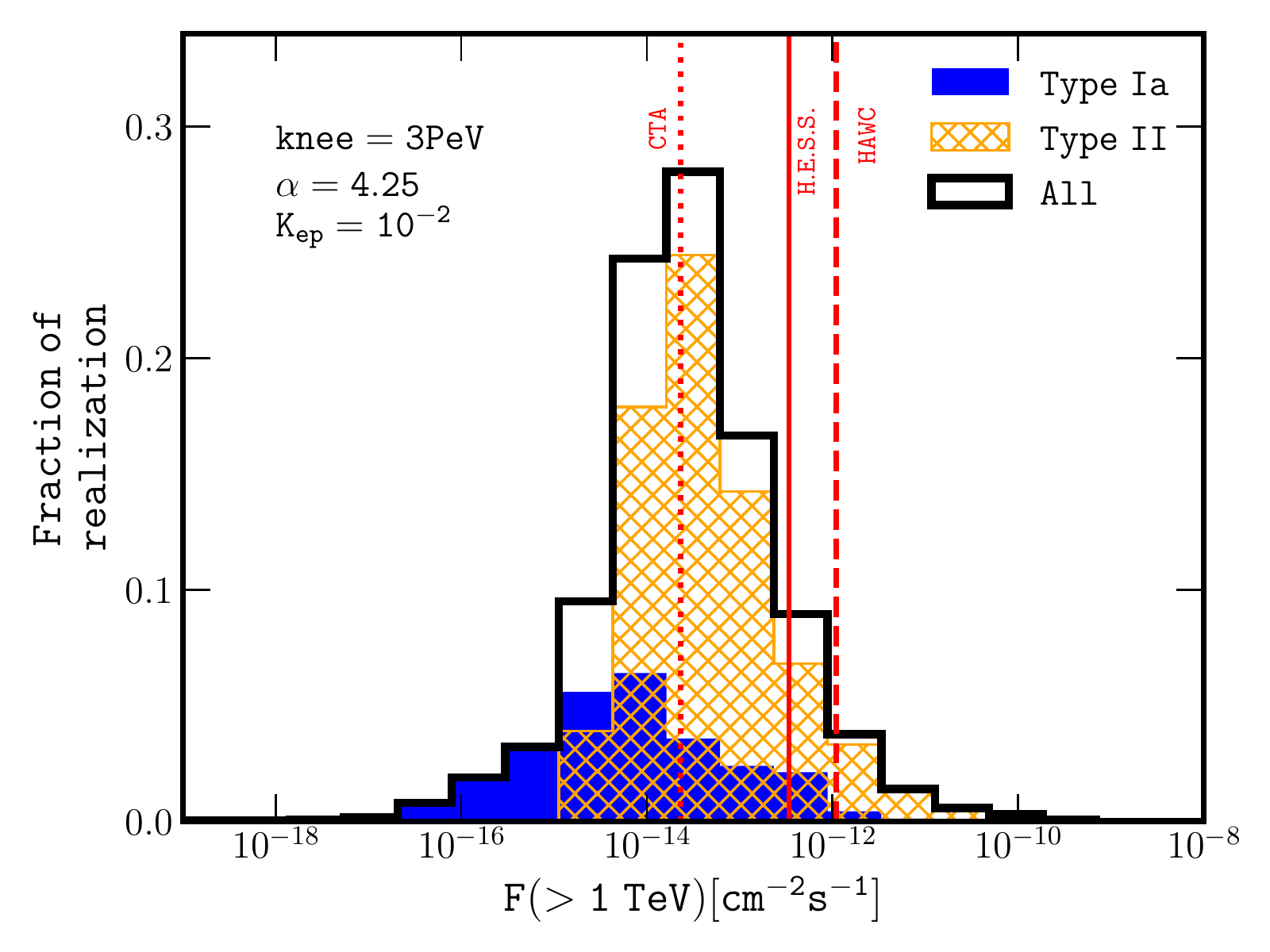}
\endminipage\hfill
\minipage{0.48\textwidth}
\includegraphics[scale=0.56]{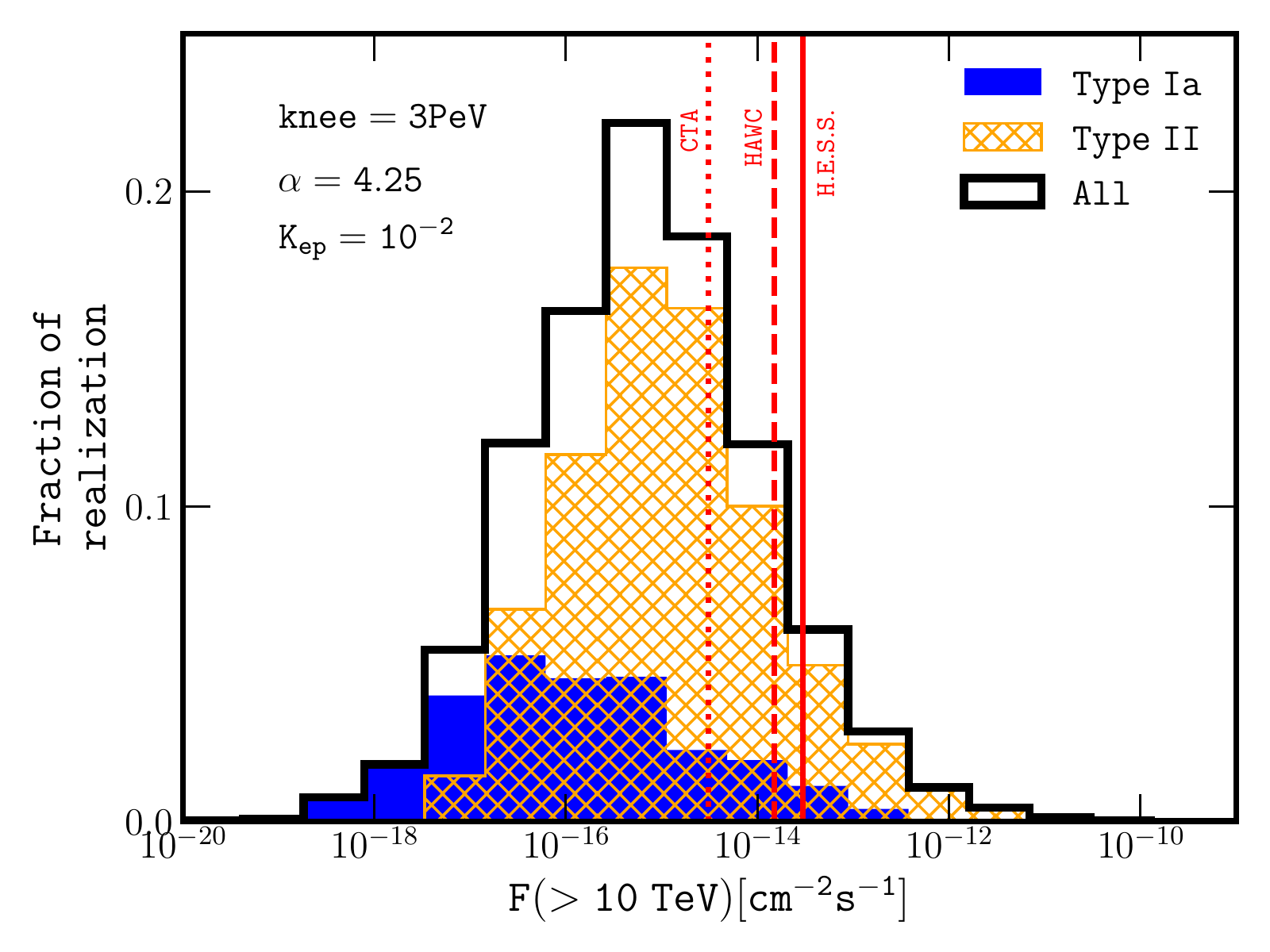}
\endminipage
\caption{Integral gamma--ray flux  of simulated pevatrons for photons of energy greater than 1 TeV (left panel) and 10 TeV (right panel). The populations from Type Ia and Type II progenitors are represented in blue (filled) and yellow (hatched), respectively. The black (solid) curve corresponds to the sum. Vertical lines correspond to typical sensitivities of H.E.S.S (solid), HAWC (dashed) and CTA (dotted) achieved during their respective Galactic plane surveys.}
\label{fig:histoflux}
\end{center}
\end{figure*}

Let us now investigate the observable properties of a typical pevatron candidate:~the integrated flux, the slope of the gamma--ray spectrum and the angular size of the typical sources. 
In Fig.~\ref{fig:histoflux} are plotted the distributions of integral fluxes of PeVatrons for photons of energy greater than 1~TeV and 10~TeV (left and right panel, respectively). For these histograms, we consider a slope $\alpha =4.25$ and $K_{\rm ep}=10^{-2}$, and we represent the populations from Type Ia  progenitors (blue filled curve),  Type II progenitors (yellow hatched curve), and the sum of both (black solid line).The vertical red lines represent the typical sensitivities reached by H.E.S.S. (GPS, \citealt{donath2016}), HAWC (5-years \citealt{HAWC2013}), and CTA (mCrab level, expected for the survey of the Galactic plane, \citealt{scienceCTA}).
The distribution is centered around a median value F$(>1$~TeV) $\approx 5 \times 10^{-13}$~cm$^{-2}$~s$^{-1}$ for the Type II progenitor and F$(>1$~TeV) $\approx 5 \times 10^{-15}$~cm$^{-2}$~s$^{-1}$ for Type Ia, and F$(>1$~TeV) $\approx 5 \times 10^{-13}$~cm$^{-2}$~s$^{-1}$ for the entire population. Current H.E.S.S.--like TeV instruments can reach a typical sensitivity of 1\% of the crab, $\approx 2-3  \;10^{-13}$~cm$^{-2}$~s$^{-1}$ therefore roughly suggesting they could potentially detect above 1 TeV a fourth of the entire pevatron population, mostly from core--collapse progenitors.
The problem here is that, for values of $K_{ep}$ not much smaller than $\approx 10^{-2}$, the leptonic emission is expected to contribute as much as the hadronic one to the total gamma-ray emission. However, as shown in Fig.~\ref{fig:spectra}, the origin of the gamma-ray emission is unambiguously hadronic if we consider photon energies above $\sim 10$ TeV.

For this reason, in the right panel of Fig.~\ref{fig:histoflux} we show the distribution of integral fluxes of pevatrons for photon energies above 10 TeV. The two types of supernovae account for two maxima centered around F$(>10$~TeV) $\approx 10^{-16}$~cm$^{-2}$~s$^{-1}$ and F$(>10$~TeV) $\approx 10^{-14}$~cm$^{-2}$~s$^{-1}$ for Type Ia and Type II progenitors, respectively. Typical sensitivity of H.E.S.S reaches $\approx 3 \times 10^{-14}$~cm$^{-2}$~s$^{-1}$, suggesting, as above 1 TeV, a potential detection of a fourth of the simulated pevatrons.

Remarkably, one can see from Fig.~\ref{fig:histoflux} that the detection potential for HAWC is better than that of H.E.S.S. at 10 TeV, and that the sensitivity of CTA is large enough to probe a third of the pevatron population.

\begin{figure*}
\begin{center}
\minipage{0.48\textwidth}
 \includegraphics[scale=0.56]{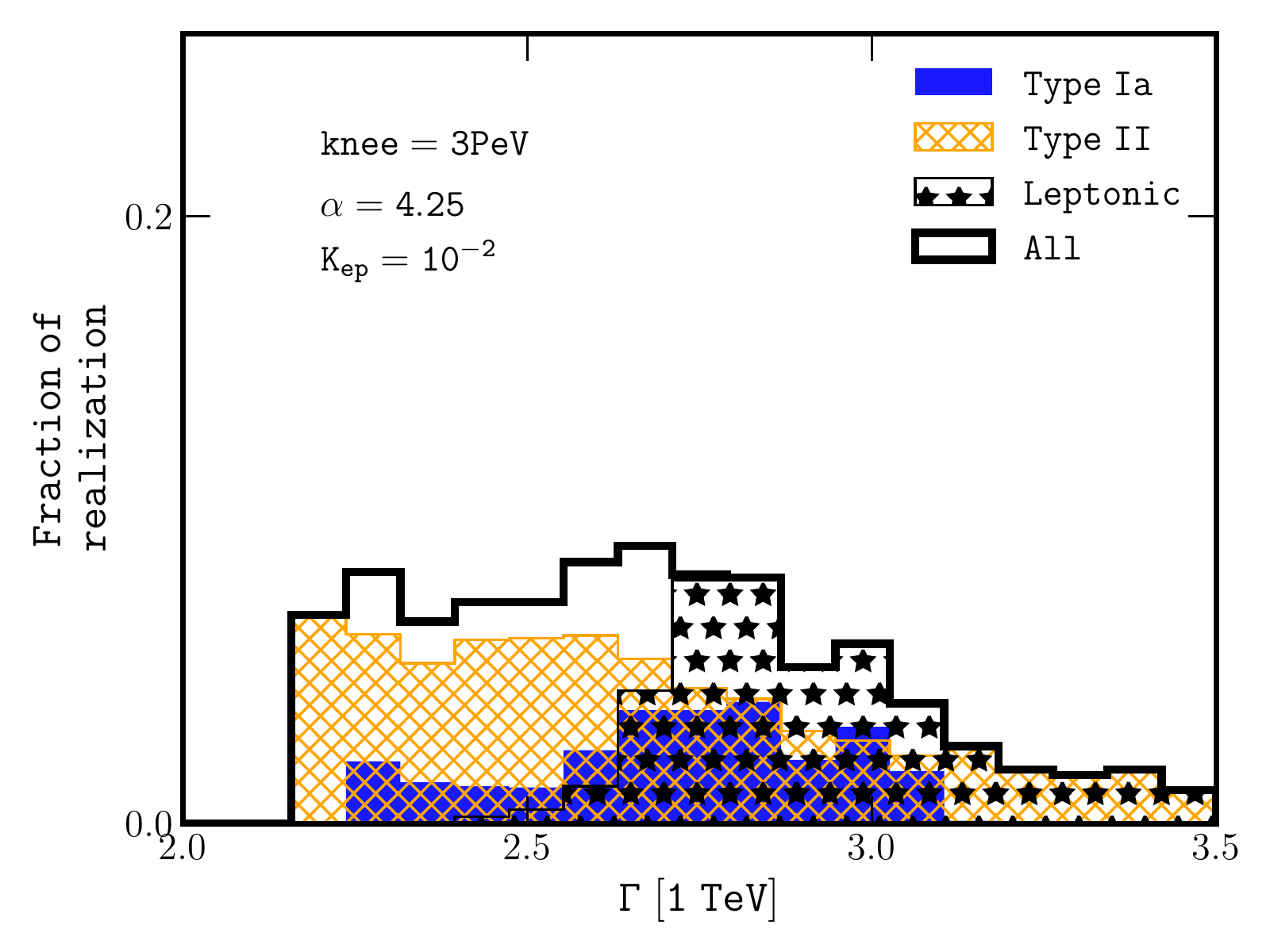}
\endminipage\hfill
\minipage{0.48\textwidth}
\includegraphics[scale=0.56]{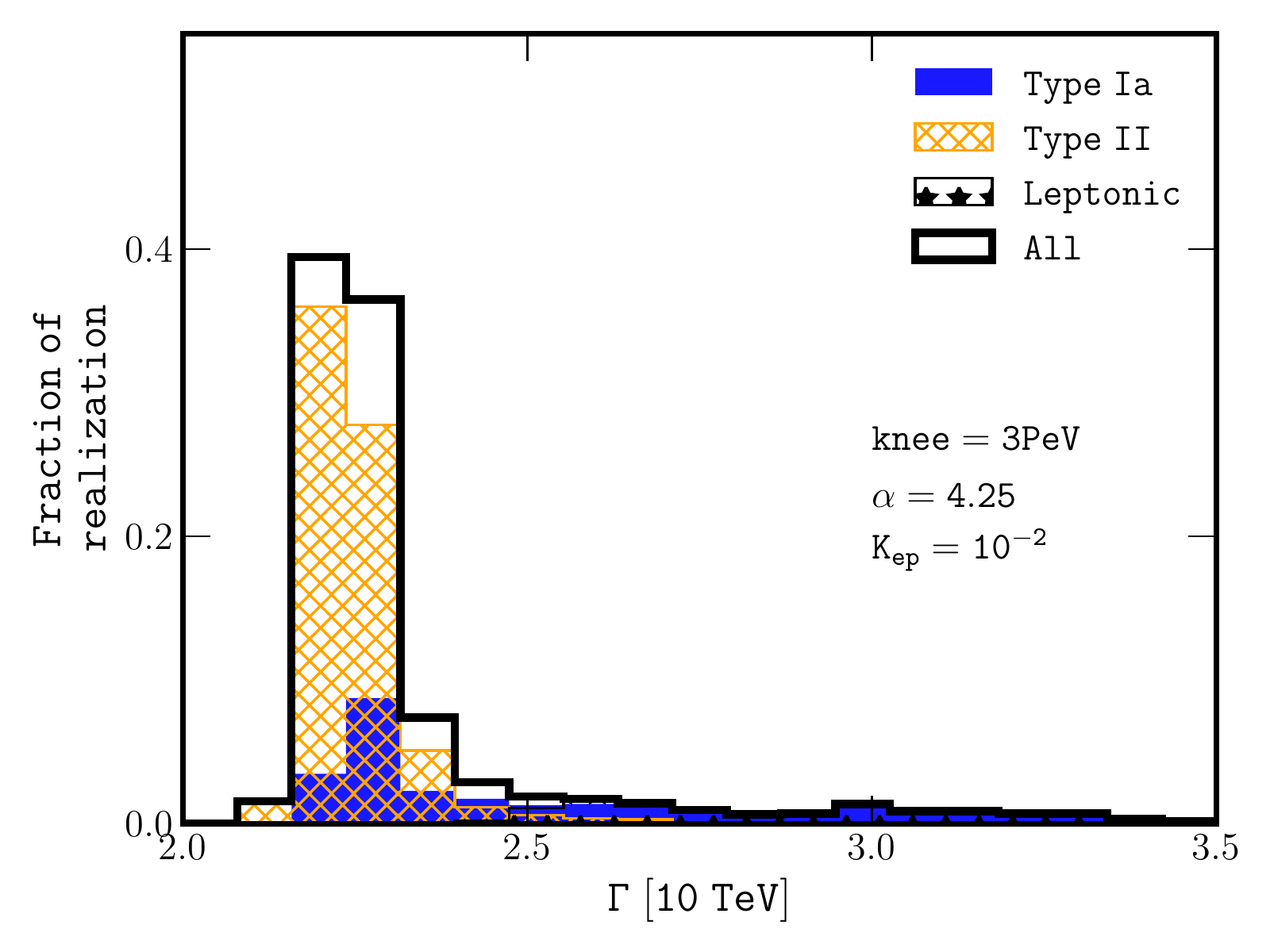}
\endminipage
\caption{Gamma--ray spectral indices  of simulated pevatrons for photons of energy greater than 1 TeV (left panel) and 10 TeV (right panel). The populations from Type Ia and Type II progenitors are represented in blue (filled) and yellow (hatched), respectively. The star shaded area corresponds to the leptonic  The black (solid) curve corresponds to the sum.}
\label{fig:histoslope}
\end{center}
\end{figure*}

We now comment on the  spectral indices of the differential gamma--ray spectra of the simulated pevatrons. Fig.~\ref{fig:histoslope} shows the distribution of indices at 1 TeV (left panel) and 10 TeV (right panel), for $K_{ep} = 10^{-2}$. 
The population from Type II progenitors and hadronic dominated emission accounts for the harder spectra with median indices  $\Gamma (1 \; \text{TeV}) \approx 2.3-2.4$ and  $\Gamma (10 \; \text{TeV}) \approx 2.3$. Type Ia progenitors lead to steeper spectra with median values $\Gamma (1 \; \text{TeV}) \approx 2.6$ and  $\Gamma (10 \; \text{TeV}) \approx 2.7$ and more distributed values. 
The pevatrons whose gamma--ray emission is dominated by leptonic mechanisms are mostly from Type Ia progenitors with a median index $\Gamma (1 \; \text{TeV}) = 2.6$. 
At photon energies of 10 TeV the histogram is very peaked because the emission is in the large majority of cases purely hadronic and thus the slope of the gamma-ray spectrum reflects the slope of the parent proton spectrum. The plot has been produced under the assumption that all SNRs accelerate proton spectra of identical slope $\alpha = 4.25$. A dispersion in the value of $\alpha$ would obviously result in a broadening of the histograms.

Another important observable is the angular extension of the gamma-ray emission. For the scenario considered in this paper, the simulated pevatrons are all quite young ($\sim$ few hundred years old) SNRs. As a consequence, the radius of the SNR shell is expected to be not very large (pc scale). The median apparent angular size of the detectable pevatrons is found to be equal to 0.8 arcmin, and over 95\% of sources are under 6 arcmin. In the TeV range, this corresponds to the typical PSF of current instruments, and therefore virtually all of the potentially detectable SNR PeVatrons will be point--like, or very marginally extended. 

To conclude, we comment on the cases of two young and very well--studied SNRs detected in the TeV range:~Cassioppea~A~\citep{casA,krause2008,kumar2015} and Tycho~\citep{tycho,tycho2}, of age $\approx 450$ years and $\approx 350$ years. Recent observations of these SNRs seem to suggest that they are not pevatrons at the moment. In the case of Tycho, from a Type Ia progenitor this is in fact quite in agreement with our results where only $\approx 0.10$ of the pevatrons are from thermonuclear supernovae. In the case of Cassioppea~A, this could be because of the very peculiar Type of the progenitor (type IIb). This rare Type of progenitor, supposed to account for less than  4$\%$ of all supernovae~\citep{seo}, releases a remarkably large total explosion energy of the order of several times $10^{51}$~erg and is surrounded by an extremely dense wind. The peculiarity of such object might explain why Cas A does not seem to fit with the general scenario described here,
where such objects are not taken into account. 
In other words, our results can be seen as compatible with the fact that these SNRs are not active pevatrons at the present time.

\section{Conclusions}
\label{sec:conclusions}

We have investigated the characteristics of the pevatron population using simulations of the Galactic population of SNRs.
As a working hypopthesis, we assumed that Galactic SNRs are the sources of Galactic CRs up to PeV particle energies. In order to reproduce the observed position of the CR proton knee in the CR spectrum, the acceleration of PeV particles has to take place at the transition between the free-expansion and Sedov phases of the SNR evoluiton. We then used Monte Carlo methods to simulate the time and location of SNRs in the Galaxy, and compute their gamma--ray emission in the TeV and multi-TeV energy range. 

In the most optimistic case, i.e. $E_{knee} = 3$ PeV, the mean number of expected detections for the typical sensitivity of current instruments operating in the TeV range (we adopted here the sensitivity of H.E.S.S.) are found to be in the range $\lesssim 10$ for the whole sky, the exact value depending on the value of the slope of accelerated particles assumed at the shock. The value of $\approx 10$ detection is obtained for a hard spectrum of accelerated particles equal to $\alpha = 4.1$, while for spectra steeper than $\alpha = 4.3$ one could expect either few detections or no detections, depending on the exact (and a priori unknown) location of SNR pevatrons in the Galaxy.
These numbers are roughly reduced by a factor of $\approx$2 if we restrict our analysis to the region of sky covered by the GPS of H.E.S.S.~$ 260^{\circ}<  l < 70^{\circ}$, $| b | < 3^{\circ}$. 
In a more pessimistic scenario where $E_{knee} = 1$ PeV the expected number of detections is at most a few, and compatible within one standard deviation with no detections. 




The unambiguous smoking gun of proton acceleration up to the PeV domain is the detection of an unattenuated gamma-ray emission extending up to the multi-TeV domain \citep[e.g.][]{gabici2007}. The presence or absence of a cutoff in the multi-TeV gamma ray spectrum can be revealed only for bright sources, such as the one recently detected in the Galactic centre region \citep{HESSpevatron}. In most cases, the gamma-ray fluxes predicted in this paper for SNR pevatrons are not bright enough to allow to identify such objects as PeV particle accelerators.   
However, the study presented here demonstrates that, at least in the most optimistic scenario, current instruments might have already detectet several SNR pevatrons. Future instruments, in particular the Cherenkov Telescope Array, will increase the number of such detections and will provide us with high quality spectra in the multi-TeV domain which are the essential observable in order to identify CR pevatrons. We showed in this paper that, if SNRs are the sources of Galactic CRs up to the knee, the search for PeV particle acceleration should be concentrated onto point-like or marginally extended sources showing an indication for a hard spectrum at $\gtrsim 10$ TeV photon energies. As associated synchrotron X-ray flux from secondary electrons must also be present, and provides a lower limit on the observed X-ray emission from SNR pevatrons at the level of $\gtrsim 10^{-14}$ erg/cm$^2$/s for objects whose gamma-ray flux is large enough to be detected by H.E.S.S.

In a future study, we aim at confronting the results of our simulations with the actual catalog of known SNRs~\citep{veritas2015,FermiSNRcatalog,HAWC2017,HESSSNRcatalog}, also considering the objects not detected in the very--high--energy  range, and considering the unidentified objects, in order to identify promising pevatron candidates  and orient future observations.

\section*{Acknowledgments}
SG and RT acknowledge support from Agence Nationale de la Recherche (grant ANR- 17-CE31-0014) and from the Observatory of Paris (Action F\'ed\'eratrice CTA). TBH  acknowledges the generous support of the National Science Foundation under cooperative agreement PHY-1352567. PC acknowledges support from the Frontiers of Science fellowship at Columbia University.


\begin{thebibliography}{199}

\bibitem[H.~E.~S.~S.~Collaboration et al.(2018)]{HESSSNRcatalog} H.~E.~S.~S.~Collaboration, :, Abdalla, H., et al.\ 2018, arXiv:1802.05172 

\bibitem[Abeysekara et al.(2013)]{HAWC2013} Abeysekara, A.~U., Alfaro, R., Alvarez, C., et al.\ 2013, Astroparticle Physics, 50, 26 

\bibitem[Abeysekara et al.(2017)]{HAWC2017} Abeysekara, A.~U., Albert, A., Alfaro, R., et al.\ 2017, ApJ, 843, 40 






\bibitem[HESS Collaboration et al.(2016)]{HESSpevatron} 
HESS Collaboration, Abramowski, A., Aharonian, F., et al.\ 2016, Nat, 531, 476 

\bibitem[Acciari et al.(2011)]{tycho}
Acciari, V.~A., Aliu, E., Arlen, T., et al.\ 2011, ApJL, 730, L20 
\bibitem[Acero et al.(2016)]{FermiSNRcatalog}
Acero, F., Ackermann, M., Ajello, M., et al.\ 2016, ApJs, 224, 8 






\bibitem[Aharonian et al.(2006)]{2006crab} 
Aharonian, et al.\ 2006, Aap, 457, 899 

\bibitem[Ahnen et al.(2017)]{casA}
Ahnen, M.~L., Ansoldi, S., Antonelli, L.~A., et al.\ 2017, MNRAS, 472, 2956 



\bibitem[Antoni et al.(2005)]{antoni2005}
 Antoni, T., Apel, W.~D., Badea, A.~F., et al.\ 2005, Astroparticle Physics, 24, 1 

\bibitem[Archambault et al.(2017)]{tycho2} Archambault, S., Archer, A., Benbow, W., et al.\ 2017, ApJ, 836, 23 




\bibitem[Bartoli et al.(2015)]{bartoli2015}
Bartoli, B., et al. 2015, Phys. Rev. D, 92, 092005

\bibitem[Bell(1978)]{bell1978} 
Bell, A.~R.\ 1978, MNRAS, 182, 147

\bibitem[Bell(2004)]{bell04}
Bell, A.~R., 2004, MNRAS, 353, 550

\bibitem[Bisnovatyi--Kogan \& Silich(1995)]{bisnovati}
Bisnovatyi--Kogan, G.~S., Silich, S.~A., 1995, Rev. Mod. Phys., 67, 661


\bibitem[Blandford \& Ostriker(1978)]{ostriker1978} 
Blandford, R.~D., \& Ostriker, J.~P.\ 1978, ApJl, 221, L29 


\bibitem[Blumenthal \& Gould(1970)]{gould}
Blumenthal, G.~R., Gould, R.~J., 1970, Rev. Mod. Phys., 42, 237

\bibitem[Chevalier(1982)]{chevalier}
Chevalier, R.~A., 1982, ApJ, 258, 790

\bibitem[Cristofari et al.(2013)]{cristofari1}
Cristofari, P., Gabici, S., Casanova, S., Terrier, R., Parizot, E., 2013, MNRAS, 424, 2748

\bibitem[Cristofari et al.(2017)]{cristofari2}
Cristofari, P., Gabici, S., Humensky, T.~B., Santander, M., Terrier, R., Parizot, E., Casanaova, S., 2017, MNRAS, 471, 201

\bibitem[CTA consortium (2013)]{CTA2013}
CTA Consortium, Acharya, B. S., Actis, M. et al., 2013, Astropart. Phys., 43, 3

\bibitem[Cherenkov Telescope Array Consortium et al.(2017)]{scienceCTA} Cherenkov Telescope Array Consortium, T., :, Acharya, B.~S., et al.\ 2017, arXiv:1709.07997 




\bibitem[Donath et al.(2017)]{donath2016} 
Donath, A., Brun, F., Chaves, R.~C.~G., et al.\ 2017,  6th International Symposium on High Energy Gamma-Ray Astronomy, 1792, 040001 

\bibitem[Drury \& Downes(2012)]{drury2012}
Drury, L.O'C., Downes, T.P.\ 2012, MNRAS, 427, 2308

\bibitem[Gabici \& Aharonian(2007)]{gabici2007}
Gabici, S., Aharonian, F.A.\ 2007, ApJ, 665, L131 

\bibitem[Gaggero et al.(2017)]{gaggero2017} 
Gaggero, D., Grasso, D., Marinelli, A., Taoso, M., \& Urbano, A.\ 2017, arXiv:1702.01124 

\bibitem[Gaggero et al.(2018)]{gaggero2018}
Gaggero, D., Zandanel, F., Cristofari, P., \& Gabici, S.\ 2018, MNRAS, 475, 5237 




\bibitem[Gabici et al.(2016)]{gabici2016}
Gabici, S., Gaggero, D., \& Zandanel, F.\ 2016, arXiv:1610.07638 

\bibitem[Giacalone \& Jokipii(2007)]{giacalone2007}
Giacalone, J., Jokipii, J.R.\ 2007, ApJ, 663, L41

\bibitem[Jouvin et al.(2017)]{terrier2017}
 Jouvin, L., Lemi{\`e}re, A., \& Terrier, R.\ 2017, MNRAS, 467, 4622 

\bibitem[Kelner et al.(2006)]{kelner}
Kelner, S.~R., Aharonian, F.~A., Bugayov, V.~V., 2006, Phys. Rev. D, 74, 034018


\bibitem[Krause et al.(2008)]{krause2008}
Krause, O., Birkmann, S.~M., Usuda, T., et al.\ 2008, Science, 320, 1195 

\bibitem[Kumar et al.(2015)]{kumar2015}
Kumar, S., \& for the VERITAS Collaboration 2015, arXiv:1508.07453 



\bibitem[Li et al.(2011)]{li2011}
Li, W., Chornok, R., Leaman, J., Filippenko, A.~V., Poznanski, D., Wang, X., Ganeshalingam, M., Mannucci, F., MNRAS, 2011, 412, 1473


\bibitem[Longair(1994)]{longair} 
Longair, M.~S.\ 1994, High energy astrophysics.~Volume 2.~Stars, the Galaxy and the interstellar medium., by Longair, M.~S..~ Cambridge University Press, Cambridge (UK), 1994, 410 p., ISBN 0-521-43439-4


\bibitem[Lorimer(2004)]{lorimer}
Lorimer, D.~R., 2004, In: F. Camilo \& B.~M. Gaensler (eds.) IAU Symposium no. 218, p. 105 


\bibitem[Nakanishi \& Sofue(2003)]{H1}
Nakanishi, H., \& Sofue, Y.\ 2003, Pasj, 55, 191 

\bibitem[Nakanishi \& Sofue(2006)]{H2} Nakanishi, H., \& Sofue, Y.\ 2006, Pasj, 58, 847 








\bibitem[Ostriker \& McKee(1995)]{ostriker}
Ostriker, J.~P., McKee, C.~F., 1988, Rev. Mod. Phys., 60, 1

\bibitem[Popkow et al.(2015)]{veritas2015}
Popkow, A., \& VERITAS Collaboration 2015, 34th International Cosmic Ray Conference (ICRC2015), 34, 750 




\bibitem[Ptuskin \& Zirakashvili(2005)]{pz05} 
Ptuskin, V.~S., Zirakashvili, V.~N., 2005, A\&A, 429, 755

\bibitem[Ptuskin et al.(2010)]{seo}
Ptuskin, V., Zirakashvili, V., Seo, E.-S., 2010, ApJ, 718, 31

\bibitem[Schure \& Bell(2013)]{schure2013} Schure, K.~M., \& Bell, A.~R.\ 2013, MNRAS, 435, 1174 



\bibitem[Weaver et al.(1977)]{weaver1977}
Weaver, R., McCray, R., Castor, J., Shapiro, P., \& Moore, R.\ 1977, ApJ, 218, 377 


\bibitem[Yusifov \& K\"u\c{c}\"uk(2004)]{PSR}
Yusifov, I, K\"u\c{c}\"uk, I, 2004, A\&A, 422, 545


\end{thebibliography}
\end{document}